\documentclass[twocolumn,showpacs,preprintnumbers,amsmath,amssymb,pra]{revtex4}
\usepackage{graphicx}

\newcommand{\vv}{\widetilde{v}}
\newcommand{\uu}{\widetilde{u}}
\newcommand{\x}{\mathbf{x}}
\newcommand{\be}{\begin{equation}}
\newcommand{\ee}{\end{equation}}
\newcommand{\bea}{\vspace{0.25cm}\begin{eqnarray}}
\newcommand{\eea}{\end{eqnarray}}

\def\PRL{{Phys. Rev. Lett.} }
\def\PRA{{Phys. Rev.} A }

\begin{document}

\title{ The possibility of absolute calibration
of analog detectors by using parametric down-conversion: a systematical
study}

\author{Giorgio Brida$^1$, Maria Chekhova$^2$, Marco Genovese$^1$, Alexander
Penin$^2$, Ivano Ruo-Berchera$^1$}

\affiliation{ $^1$ Istituto Elettrotecnico Nazionale Galileo
Ferraris, Strada delle Cacce 91, 10135 Torino, Italy \\ $^2$
Physics Department, M.V. Lomonosov State University, 119992
Moscow, Russia. }

\begin{abstract}
In this paper we systematically study the possibility of the
absolute calibration of analog photo-detectors based on the
properties of  parametric amplifiers. Our results show that such a
method can be effectively developed with interesting possible
metrological applications.
\end{abstract}


\maketitle

\section{  Introduction}

Accurate calibration of photodetectors both in analog and in
photon-counting regime is fundamental for various scientific
applications, which range  from "traditional" quantum optics
\cite{QO} to the studies on foundations of quantum mechanics
\cite{au}, quantum cryptography \cite{QCr}, quantum computation
\cite{QCo}, etc.

In traditional optical radiometry primary standards are based on
absolute sources or detectors \cite{anal}. Synchrotron and
blackbody radiations are absolute sources. The spectral radiance
of radiation emitted by electron storage rings can be predicted
using the Schwinger equation provided that the magnetic field and
the electron current have been measured. In a similar way, the
radiant flux emitted by a blackbody at a known thermodynamic
temperature can be calculated from Planck's equation. Both sources
provide radiation with continuous spectral distribution, the
dominant contribution being in the soft X-ray and UV spectral
region for synchrotron radiation and in the IR spectral region for
blackbody radiation. The relative uncertainty of both these
sources is about 1 part in $10^{3}$.  Among the absolute
detectors, there exist the following two types: thermal detector,
called electrical substitution radiometer (ESR), and semiconductor
photon detector. The ESR is based on the electrical substitution
principle, that is, the heating effect of the unknown optical
radiation is compared with the heating effect produced by a
measured quantity of electrical power (Joule effect). The
operation of these detectors at cryogenic temperatures
\cite{analcrio}, below 20 K, allows to reduce measurement
uncertainty  down to 1 part in $10^{4}$. The operation principle
of semiconductor photon detectors underpins on the photoelectric
effect, that is, the generation of free electron-hole pairs at the
absorption of a photon. The quantum efficiency is defined as the
average number of free electron-hole pairs produced per incident
photon. In a high-quality silicon photodiode the quantum
efficiency  for the visible  range is close to unity to within a
few tenths of one percent. The deviation of the quantum efficiency
from unity could be determined, independently from other
radiometric measurements, by self-calibration technique
\cite{analtrap}. Uncertainties of a few parts in $10^{4}$ appear
to be the limit of this technique with commercial photodiodes.
There is a perspective of improvement to 1 part in $10^{6}$ or
better with custom photodiodes operated at liquid nitrogen
temperatures \cite{Geist}

For what concerns single-photon detectors, classical calibration
techniques are based on the use of a strongly attenuated laser
source whose (unattenuated) intensity has been measured by means
of a power-meter. The uncertainty of this kind of measurements is
principally limited by the uncertainty in the calibration of the
very low transmittance required for reaching single-photon level.

This limitation has initiated the study of an alternative scheme,
based on the use of photons produced by means of spontaneous
parametric down-conversion (SPDC), where photons are emitted in
pairs strongly correlated in direction, wavelength and
polarization. Furthermore, photons of the same pair are emitted
within tens of femtoseconds. Since the observation of a photon of
a pair on a certain direction (signal) implies the presence of
another one on the conjugated direction (idler), when this last is
not observed this occurs because  of the non-ideal quantum
efficiency of the idler detector, which can be measured in this
way \cite{bp1,klysh,malygin,KP,alan}. This absolute technique (and
others related \cite{poc}) is becoming attractive for national
metrological institutes to establish absolute radiometric
standards because it relies simply on the counting of events,
involves a remarkably small number of measured quantities, and
does not require any standards.

Because of the success of the SPDC scheme for calibrating single-photon
detectors,
it is important to analyse if similar absolute calibration methods
could be developed for analog ones, eventually allowing the
development of a calibration scheme operating in both regimes.

A seminal attempt in this sense was done in \cite{SP} following
the theoretical proposal of \cite{klysh}. Nevertheless, these
results were limited to the case of very low intensity (as we will
show in detail later) and were very far from being developed to
the metrological level. An accurate analysis of the possibility to
calibrate analog detectors by using SPDC overcoming these limits
is therefore demanded.

Incidentally, the quantum efficiency $\eta$ of analog detectors
appears also in equations describing suppression of photon noise
in parametric down-conversion using the feedforward \cite{Mertz}
or feedback \cite{Shapiro},\cite{Tapster} transformations. Hence,
such experiments could eventually be used to develop an
alternative scheme for analog detector calibration.

The purpose of this paper is  a systematical study of the
possibility to calibrate analog detectors by using parametric down
conversion.

In section II we will give a short presentation of the SPDC scheme
used in the photon-counting regime. In Section III we will
analyze multimode SPDC, following a theoretical description of
SPDC \cite{BGBL} developed with an account for the high-gain
regime (of which we give a short summary). After discussing some
general results, we suggest possible calibration methods and show
how the scheme of Refs. \cite{klysh,SP} can be derived as the
low-intensity limit of one of them.  Finally, in Section IV
we will consider the possibility of calibrating analog detectors
using the schemes for photocurrent fluctuation suppression.
In particular, two-mode squeezing and
feedforward techniques will be discussed. For these schemes, we
will analyse new possibilities and limitations.

\section{ The SPDC scheme for calibrating single-photon detectors}

The scheme for calibrating single-photon detectors by using SPDC
is based on the specific properties of this process, where a
photon of the pump beam (usually a laser beam) "decays" inside a
non-linear crystal in two lower-frequency photons, 1 and 2
(conventionally dubbed "idler" and "signal"), such that energy and
momentum are conserved ($\omega_{pump} = \omega_{1} + \omega_{2}$,
$\vec{k}_{pump} = \vec{k}_{1} + \vec{k}_{2}$). Moreover, the
two photons are emitted within few femtoseconds. In synthesis, the
calibration procedure consists \cite{KP} of placing a couple of
photon-counting detectors down-stream to the nonlinear crystal,
along the direction of propagation of correlated photon pairs for
a selected pair of frequencies: the detection of an event by one of the
two detectors guarantees with certainty, due to the SPDC
properties, the presence of a photon with a fixed wavelength
on the conjugated direction. If  $N$ is the total number of photon
pairs emitted from the crystal in a given time interval
and $\langle N_{1}\rangle$, $\langle N_{2}\rangle$ and $\langle
N_{c}\rangle$ are the mean numbers of events recorded during the
same time interval
by the signal detector, the idler detector,
and in coincidence, respectively, we have the following obvious
relationships \cite{klysh}:

\be \langle N_{1}\rangle = \eta_{ 1} N ; \, \, \langle
N_{2}\rangle= \eta_{ 2} · N, \label{1} \ee
where $\eta_{1}$ and $ \eta_{2}$ are the detection efficiencies in
the signal and idler arms. The number of coincidences is

\be\label{2} \langle N_{c}\rangle = \eta_{ 1} \eta_{2}·N,  \ee
due to the statistical independence of the two detectors. Then the
detection efficiency can be found as

\be \label{3}\eta_{ 1} = \langle N_{c}\rangle / \langle
N_{2}\rangle. \ee

This simple relation, slightly modified by taking into account the
background subtraction and corrections for the acquisition
dead-time, is the basis for the scheme for absolute calibration of
single-photon detectors by means of SPDC, which reaches now
measurement precision competitive  with traditional methods
\cite{alan}.
  \vskip 1cm

\section{Analog detection} \vskip 0.5cm
\subsection{Basic formulas}

In order to study the possibility of absolute calibration of
analog detectors we are interested in a model of SPDC working at
any values of parametric gain. Firstly, the reason is the
necessity to work with quite large intensities yielding continuous
photocurrent. Secondly, we need to explore new possibilities to
use the properties of SPDC for calibration without the usual
low-gain limitation. A theory suited for these purposes is
developed in~\cite{BGBL} and literature cited therein. For
simplicity of description we will consider type-I SPDC in the
degenerate case where the signal and idler frequencies are
$\omega_{1}= \omega_{2}=\frac{\omega_{pump}}{2}$. In the limit of
monochromatic and plane-wave pump approximation, only pairs of
modes with opposite transverse wave vectors, $\mathbf{q}$ and
$-\mathbf{q}$, and with frequencies $\omega_{1}-\Omega$ and
$\omega_{2}+\Omega$ are coupled as a consequence of energy and
transverse momentum conservation. We can write the input-output
transformation relating the field operator $a(q,\Omega)$ at the
input face of the nonlinear crystal to the field operator
$b(q,\Omega)$ at the output face:

\begin{equation}\label{in-out}
  b(q,\Omega)=u(q,\Omega)a(q,\Omega)+v(q,\Omega)a^{\dag }(-q,-\Omega).
\end{equation}

The coefficients $u$ and $v$ are  considered in \cite{BGLK}. For our
analysis, we are interested not in the exact form of $u$ and
$v$, but rather in the properties

\begin{eqnarray}
|u(\mathbf{q},\Omega)|^{2}-|v(\mathbf{q},\Omega)|^{2} =  1,    \nonumber  \\
u(\mathbf{q},\Omega)v(-\mathbf{q},-\Omega)=u(-\mathbf{q},-\Omega)v(\mathbf{q},\Omega),
\\
\end{eqnarray}
which guarantee the conservation of the free-field commutation
relations
$[b(\mathbf{q},\Omega),b^{\dag}(\mathbf{q},\Omega)]=\delta(\mathbf{q}-\mathbf{q}')\delta(\Omega-\Omega')$
and $[b(\mathbf{q},\Omega),b(\mathbf{q},\Omega)]=0$.

The far field  observed in the focal plane of a thin lens of focal
length $f$ is obtained in \cite{BGLK} from the near field by means
of the following transformation:

\begin{equation}\label{far field operator}
B(\mathbf{x},t)=\frac{-i}{\lambda_{s}f}\int_{S_{A}} d\mathbf{x}'
  b(\mathbf{x}',t)e^{-i\frac{2\pi}{\lambda_{s}f}\mathbf{x}\cdot\mathbf{x}'},
\end{equation}
where $\lambda_{s}=\frac{2\pi c}{\omega_{s}}$ is the central
free-space wavelength of the down-converted light and $S_{A}$ is
the transverse area of the domain where PDC takes place.
In  practical situations it can be identified
with the effective cross-section area of the pump beam. We stress
that Eq. (\ref{far field operator}) is correctly usable only for
the calculation of normally-ordered correlation functions because
it does not conserve the correct commutation relations. According
to (\ref{in-out}),  (\ref{far field operator}), one can write the
far field as

\begin{eqnarray}\label{B(x,t)}
B(\mathbf{x},t)=\frac{2\pi
i}{\lambda_{s}f}\int\frac{d\Omega}{\sqrt{2\pi}}e^{-i\Omega t}\int
d\mathbf{x}'p(\mathbf{x}-\mathbf{x}')\cdot \hspace{1cm} \nonumber\\
\cdot[\widetilde{u}(\mathbf{x}',\Omega)a(\frac{2\pi}{\lambda_{s}f}\mathbf{x}',\Omega)+
\widetilde{v}(\mathbf{x}',\Omega)a^{\dag }(-\frac{2\pi}{\lambda_{s}f}\mathbf{x}',-\Omega)],\\
\end{eqnarray}
where
\begin{eqnarray}
\uu(\x,\Omega)=u(\frac{2\pi}{\lambda_{s}f}\x,\Omega),\nonumber\\
\vv(\x,\Omega)=v(\frac{2\pi}{\lambda_{s}f}\x,\Omega).
\end{eqnarray}

The spatial variation scale of these coefficients is on the order
of $x_{0}=\sqrt{\frac{\lambda_{s}f}{2\pi l_{c}}}$, $l_{c}$ being
the length of the crystal. One can interpret $x_0$ as the
transverse coherence length of SPDC in the focal plane of the
lens. The variation scale of $\uu$ and $\vv$ in frequency, let us
denote it $\Omega_{0}$, represents the typical bandwidth of SPDC
in the temporal frequency domain and
$\tau_{coh}=\frac{1}{\Omega_{0}}$ is the coherence time. At the
same time,

\begin{equation}\label{p}
p(\x)=\left(\frac{-i}{\lambda_{s}f}\right)^{2}\int_{S_{A}} d\mathbf{x}'
  e^{-i\frac{2\pi}{\lambda_{s}f}\mathbf{x}\cdot\mathbf{x}'}
\end{equation}
is the diffraction pattern in the far-field plane due to the
finite transverse  size of the system. Also here we are not
interested in the exact form of $p(\x)$, which is determined by
the shape of $S_{A}$, but we observe that its typical size  is
$S_{diff}=(\lambda_{s}f)^{2}/S_{A}$ and its amplitude
$\frac{1}{S_{diff}}$. Hereafter let us assume the size of
$S_{diff}$ to be much smaller than the  coherence length of PDC in
the far field, i.e., $x_{diff}\ll x_{0}$. Within this
approximation the integrals can be evaluated considering $p(\x)$
as a delta function. Now we can calculate the mean value  of the
photon flux density operator $I(\x,t)\equiv
B^{\dag}(\mathbf{x},t)B(\mathbf{x},t)$ in the detection plane. By
using Eqs. (\ref{B(x,t)}) and (\ref{in-out}), and considering the
vacuum as the input state, one obtains

\begin{equation}\label{I(x,t)}
\langle I(\x,t)\rangle= \frac{1}{S_{diff}}\int d\Omega
|\vv(\mathbf{x},\Omega)|^{2}.
\end{equation}

The physical meaning of this quantity is the mean number of
photons crossing the detection plane at point $\x$ at time $t$ per
unit area and time. The integral function is usually referred to
as the spectral gain of SPDC and its height represents the mean
number of photons per coherence time.

We are also interested in the second-order correlation function of
the intensity fluctuations defined as
\begin{equation}\label{I(x,t)I(x',t')}
\langle \delta I(\x,t)\delta I(\x',t')\rangle\equiv\langle
I(\x,t)I(\x',t')\rangle-
\langle I(\x,t)\rangle\langle I(\x',t')\rangle,\\
\end{equation}
where $\langle I(\x,t)I(\x',t')\rangle$ is determined by the joint
probability of a photon reaching the detection plane at $\x'$ at
time $t'$ and the other one, at $\x$ at time $t$. It is convenient
to distinguish between the two contributions, one due to the
autocorrelation inside one beam (signal or idler), the other one
due to the cross-correlation between the two beams:
\begin{eqnarray}\label{I(x,t)I(x',t')2}
\langle \delta I(\x,t)\delta I(\x',t')\rangle=\mathcal{G}_{11}(\x,t,\x',t')
+\mathcal{G}_{12}(\x,t,\x',t').\nonumber\\
\end{eqnarray}
Here $\mathcal{G}_{11}(\x,t,\x',t')$ represents the
autocorrelation contribution, when the distance between $\x$ and
$\x'$ is less than $x_{diff}$. On the contrary,
$\mathcal{G}_{12}(\x,t,\x',t')$ is different from zero only if
$\x'\simeq-\x$ within $x_{diff}$. Therefore, it is the term
responsible for the well-known cross-correlation between signal
and idler beams. From Eqs. (\ref{B(x,t)}) and (\ref{in-out}), it
is easy to show that

\begin{eqnarray}\label{dI1(x,t)dI1(x',t')}
\mathcal{G}_{11}(\x,t,\x',t') = \langle
I(\x,t)\rangle\delta(\x-\x')\delta(t-t')+ \nonumber\\
+  |p(\x-\x')|^{2}\int\int \frac{d\Omega d\Omega'}{2 \pi}
[e^{-i(\Omega-\Omega')(t'-t)}\nonumber\\
\hspace{2.5cm}|\vv(\mathbf{x},\Omega)|^{2}|\vv(\mathbf{x},\Omega')|^{2}],\nonumber\\
\end{eqnarray}

\begin{eqnarray}\label{cross-corr}
\mathcal{G}_{12}(\x,t,\x',t')=|p(\x+\x')|^{2} \int\int \frac{d\Omega
d\Omega'}{2\pi} \nonumber
e^{-i(\Omega- \Omega')(t'-t)}\\
\vv^{*}(\mathbf{x},\Omega)\uu^{*}(-\mathbf{x},-\Omega)\vv(\mathbf{x},\Omega')\uu(-\mathbf{x},-\Omega').\nonumber\\
\end{eqnarray}

The first term on the right-hand side of Eq.
(\ref{dI1(x,t)dI1(x',t')}) is due to the commutator
$[B(\mathbf{x},t),
B^{\dag}(\mathbf{x}',t')]=\delta(\mathbf{x}-\mathbf{x}')\delta(t-t')$
and it is responsible for the rise of shot noise during detection,
while the second term is the normally ordered auto-correlation. In
Eq. (\ref{cross-corr}) the shot-noise term is not present because
the commutator is null, being the cross-correlation calculated for
$\x\simeq -\x'$. Let us consider two detectors, 1 and 2, having
quantum efficiencies $\eta_{1}$ and $\eta_{2}$, to register
photons crossing two arbitrary but symmetrically placed regions
$R_{1}$ and $R_{2}$ in the detection plane. Let at least one of
the two areas be much bigger than the diffraction area $S_{diff}$
\footnote{Indeed, it is sufficient to assume that only one of the
two areas, for instance, $R_{2}$ obeys this restriction when we
will work with the cross-correlation function. We also performed
the calculation under the opposite condition, but the results do
not present any particular interest here.}. In fact, from the
works on photon-counting detectors calibration it is well known
that the area of the detector under investigation should be large
enough to collect all the photons correlated with those incident
on the trigger detector surface, otherwise the quantum efficiency
is underestimated. Integration of Eq. (\ref{I(x,t)}) over $R_{1}$
gives the photon flux reaching detector 1:

\begin{equation}\label{I1}
\langle I_{1}\rangle= \frac{1}{S_{diff}}\int_{R_{1}} d\x \int
d\Omega |\vv(\mathbf{x},\Omega)|^{2}
\end{equation}

The autocorrelation and cross-correlation functions of the photon fluxes can
be obtained by integrating definition (\ref{I(x,t)I(x',t')}),
respectively, over $R_{1}\times R_{1}$ and $R_{1}\times R_{2}$ and
using Eqs. (\ref{dI1(x,t)dI1(x',t')}) and (\ref{cross-corr}):

\begin{eqnarray}\label{self-corr-time}
\langle I_{1}(t)I_{1}(t')\rangle =\langle I_{1}\rangle^{2}+ \langle
I_{1}\rangle\delta(t-t')+ \nonumber\\
+ \frac{1}{S_{diff}}\int_{R_{1}}d\x\int\int \frac{d\Omega d\Omega'}{2 \pi}
[e^{-i(\Omega-\Omega')(t'-t)}\nonumber\\
\hspace{2.5cm}|\vv(\mathbf{x},\Omega)|^{2}|\vv(\mathbf{x},\Omega')|^{2}],\nonumber\\
\end{eqnarray}

\begin{eqnarray}\label{cross-corr-time}
\langle I_{1}(t)I_{2}(t')\rangle=\langle I_{1}\rangle\langle
I_{2}\rangle+\frac{1}{S_{diff}}\int_{R_{1}}d\x \int\int \frac{d\Omega
d\Omega'}{2\pi} \nonumber\\
e^{-i(\Omega- \Omega')(t'-t)}
\vv^{*}(\mathbf{x},\Omega)\uu^{*}(-\mathbf{x},-\Omega)\vv(\mathbf{x},\Omega')\uu(-\mathbf{x},-\Omega').\nonumber\\
\end{eqnarray}

In the quantum theory approach the shot noise arises from the
commutation relations of the free-field operators, see, for
example, \cite{photon-noise}. Any losses in the channels must be
theoretically described, so that the correct commutators are
preserved until the moment of the measurement, i.e., the
photon-electron conversion inside the detector. Thus, to take into
account the losses due to the non-ideal quantum efficiency, we
consider any real detector as an ideal one preceded by a beam
splitter with the transmission coefficient equal to the quantum
efficiency $\eta$ of the real detector. We substitute the field
$B(\mathbf{x},t)$ entering the beam splitter with the transmitted
field $C(\mathbf{x},t)$ defined as \cite{QO}

\begin{equation}\label{beam-splitter}
  C(\mathbf{x},t)=\sqrt{\eta}B(\mathbf{x},t)+\sqrt{1-\eta}V(\x,t),
\end{equation}
where $V(\x,t)$ is the field operator for the second input port of
the beam splitter, which is assumed to be in the vacuum state.
From this commutator-preserving transformation, it easily turns
out how to regard sub-unity quantum efficiency: it is sufficient
to replace, in the normally ordered products, $B(\mathbf{x},t)$
with $\sqrt{\eta}B(\mathbf{x},t)$ \cite{photon-noise}.

\subsection{Feasibility of analog detectors calibration by measuring SPDC
correlations}

For our calculations, we will consider a few millimeters
non-linear crystal pumped by a CW laser. If the waist of the pump,
identifiable with the transverse cross-section of the system
$S_{A}$, is relatively large, namely on the order of millimeter or
more, the real system fits the model of SPDC discussed in the
previous section.

In any detectors the absorbtion of a single photon, i.e., any
detection event, generates an electric pulse in the current or in
the voltage having a profile $f(t)$ and a random area $q_{n}$  at
a random time of occurrence $t_{n}$. In the analog process, with a
large incident photon flux, we cannot distinguish between two
different pulses because they overlap. Rather, the information
about the statistics of light is carried by the continuous
fluctuations of the photocurrent. We express the current (or
voltage) as a superposition of many pulses \cite{Soda}:
\begin{equation*}
  i(t)=\sum_{n}q_{n}f(t-t_{n}).
\end{equation*}
In an ideal instantaneous-response photocell, all values $q_{n}$
are equal to the charge $e$ of a single electron and $f(t)\sim
\delta(t)$. For real detectors, the time constant
$\tau_{p}$  is finite, and typically $\tau_{p}\sim 1
\hbox{ns}$ or more. Also, when a photodetector is operated in the
avalanche multiplication mode, this process gives rise to an internal
current gain. The statistical nature of the multiplication process
gives an additional contribution to the current fluctuations \cite{MNUAD}.

Since the probability density of observing a photon at time $t$ is
related to the quantum mean value of the photon flux $\langle
I(t)\rangle$, we calculate the average current as
\begin{eqnarray}\label{curr}
\langle i_{1}(t)\rangle =\sum_{n}\langle
q_{1n}f(t-t_{1n})\rangle=\nonumber\\
\int dt_{1}\langle q_{1}\rangle f(t-t_{1})\langle
I_{1}(t)\rangle\:.
\end{eqnarray}
Analogously, the quantum-mechanical second-order intensity
correlation function is determined by the probability density to have a
photon detected at time $t$ and another one at time $t'$, see Eqs.
(\ref{I(x,t)I(x',t')}). Therefore the correlation functions for
the current fluctuations can be expressed as

\begin{eqnarray}\label{auto-corr-curr}
\langle i_{1}(t)i_{1}(t+\tau)\rangle =\sum_{n,m}\langle q_{1n}q_{1m}
f(t-t_{1n})f(t-t_{1m}+\tau)\rangle=\nonumber\\
\int\int dt_{1}dt'_{1} \langle q_{1} q_{1}'\rangle
f(t-t_{1})f(t-t'_{1}+\tau)\langle
I_{1}(t_{1})I_{1}(t'_{1})\rangle,\nonumber\\
\end{eqnarray}
\begin{eqnarray}\label{cross-corr-curr}
\langle i_{1}(t)i_{2}(t+\tau)\rangle =\sum_{n,m}\langle  q_{1n}q_{2m}
f(t-t_{1n})f(t-t_{2m}+\tau)\rangle=\nonumber\\
\int\int dt_{1}dt_{2} \langle q_{1} q_{2}\rangle
f(t-t_{1})f(t-t_{2}+\tau)\langle
I_{1}(t_{1})I_{2}(t_{2})\rangle,\nonumber\\
\end{eqnarray}
where the first equation is the electron current autocorrelation
function for each detector, one registering the intensity of the
signal beam and the other registering the intensity of the idler
beam, while the second one is the cross-correlation function between the
electron currents produced by the two different detectors.

According to Eqs. (\ref{curr}) and (\ref{I1}), the mean value of
the electron current (the analog of Eq. (\ref{1}) in
photon-counting regime) is equal to
\begin{equation}\label{curr-time}
  \langle i_{1}\rangle=\eta_{1}\langle q_{1}\rangle\langle
I_{1}\rangle\:.
\end{equation}
Here the quantum efficiency has been taken into account according
to formula (\ref{beam-splitter}), with the substitution $\langle
I_{1}\rangle\rightarrow\eta_{1}\langle I_{1}\rangle$. The factor
$\langle q_{1}\rangle$ is the average charge produced in a
detection event.

The current correlation functions can be obtained substituting
Eqs. (\ref{self-corr-time}) and (\ref{cross-corr-time}),
respectively, in (\ref{auto-corr-curr}) and
(\ref{cross-corr-curr}). We note that functions in the integrals
of (\ref{self-corr-time}) and (\ref{cross-corr-time}) have
significant values for $|\Omega-\Omega'|\sim\Omega_{0}$. Roughly,
this means that the correlation functions have a sinc-like
behavior in time, with the central peak width equal to the
coherence time of PDC $\tau_{coh}=\frac{1}{\Omega_{0}}\sim
10^{-13}\hbox{s}$. As mentioned above, the resolving
time of a real analog detector is finite, and in general can be
considered much larger than the SPDC coherence time. Thus any
fluctuations in the intensity of light are averaged over
$\tau_{p}$ during the detection process. So in the limit
$\tau_{p}\gg\tau_{coh}$ we have

\begin{eqnarray}\label{self-corr-curr2}
\langle i_{1}(t) i_{1}(t+\tau)\rangle=\langle
i_{1}\rangle^{2}+\eta_{1}\langle
q_{1}^{2}\rangle \mathcal{F}(\tau)\cdot\nonumber\\
\cdot\left[\langle
I_{1}\rangle+\eta_{1}\frac{1}{S_{diff}}\int_{R_{1}}d\x\int
\frac{d\Omega}{2 \pi}|\vv(\mathbf{x},\Omega)|^{4}\right]
\end{eqnarray}
and (the analog of Eq. (\ref{2}) in the photon-counting regime)
\begin{eqnarray}\label{cross- corr-curr2}
\langle i_{1}(t)i_{2}(t+\tau)\rangle=\langle i_{1}\rangle\langle
i_{2}\rangle+\eta_{1}\eta_{2}\langle q_{1}\rangle\langle
q_{2}\rangle \mathcal{F}(\tau)\cdot\nonumber\\
\cdot\left[\langle
I_{1}\rangle+\frac{1}{S_{diff}}\int_{R_{1}}d\x\int
\frac{d\Omega}{2 \pi}|\vv(\mathbf{x},\Omega)|^{4}\right],&
\end{eqnarray}
where we have defined

\begin{equation}\label{F(t)}
  \mathcal{F}(\tau)\equiv\int dt f(t)f(t+\tau).
\end{equation}

The last two equations are the fundamental tools for studying  the
problem of absolute calibration of analog detectors and thus we
are going to discuss them in detail. Despite
(\ref{self-corr-curr2}) and (\ref{cross- corr-curr2}) seem to be
quite symmetric, we observe some important differences. The
presence of $\langle I_{1}\rangle$ in the autocorrelation function
is due to the shot noise contribution and for this reason the
quantum efficiency $\eta$ enters linearly, while in the current
cross-correlation function the corresponding term is due to the high
quantum correlation between the signal and idler beams of PDC and
the quantum efficiency appears quadratically. It is equivalent to
the right-hand side of Eq. (\ref{2}) for the counting regime and
its presence is the key for absolute calibration. The second term
inside the brackets, both for auto- and cross-correlation
functions, is important only when the number of photons per
coherence time is not close to zero and so the presence of two or
more photons within that time is not negligible. In fact,
the integral term can be estimated as $v^{2}\langle I_{1}\rangle$
and can be neglected as long as $v^{2}\ll 1$, i.e., the mean
number of photons per $\tau_{coh}$ is much smaller than one.
Anyway, if the duration of the photocurrent pulse is much larger than
the coherence time, this  assumption does not prevent photodetection to
be in a strongly analog regime, because a lot of photons can be
absorbed during the pulse duration as well. The term proportional
to $\langle I_{}\rangle^{2}$ is due to the presence of more than
one photon in time $\tau_{p}$ and for that reason is more
delicate. As we can observe in Eq. (\ref{cross- corr-curr2}) it
can be neglected only if $\langle
I_{}\rangle\ll\mathcal{F}(\tau)$. Since the pulse $f(t)$ has a
height around $1/\tau_{p}$, by Eq. (\ref{F(t)}),
$max\left[\mathcal{F}(\tau)\right]=\mathcal{F}(0)\sim 1/\tau_{p}$.
So the condition becomes $\langle I_{}\rangle\tau_{p}\ll 1$, i.e.,
the number of incident photons during the resolving time of the
detector should be much less than one, i.e.  one should work in a
non-overlapping
regime. In principle, in this case one could distinguish between
different pulses of the current and work in the counting mode.

The usual definition of the quantum efficiency is the ratio
between the number of photons detected and the number of photons
incident on the detector surface. This definition is completely
suitable in the case of counting detectors and is exactly the
meaning of $\eta$ in our paper. But in the case of analog
detection, we cannot in principle distinguish between different
current pulses. Thus, according to formula (\ref{curr-time}), we
adopt the definition of analog quantum efficiency as
$\Gamma\equiv\eta\langle q\rangle=\langle i\rangle/\langle
I\rangle$, having the meaning of the electron charge produced per
single incident photon, or the ratio between the electron flux and
the photon flux. If the charge $q$ produced per photon fluctuates
it increases the current fluctuations. This explains why $\langle
q_{1}^{2}\rangle$ appears in Eq. (\ref{self-corr-curr2}) instead
of $\langle q_{1}\rangle^{2}$. In principle, the most general way
to obtain an estimation of $\eta$ working with the PDC light
intensity in the photon-counting regime is dividing the
coincidence counting rate (proportional to the cross-correlation
function) by the detector counting rate (proportional to the
intensity). This method works because in the photon-counting
regime the temporal shape of the current pulses and their width is
not important; instead, one registers a single pulse or not
registers it. This is not the case for analog detection in which
the pulse shape $f(t)$ appears in formulas through the factor
$\mathcal{F}(\tau)$. In general, we do not know this function, and
this makes the absolute calibration of analog detectors more
difficult. However, as we are going to show, under some condition
it is possible to overcome this drawback.

Let us distinguish between three different regimes: very low
intensity (I), middle intensity (II), high intensity (III).

\textbf{(I)} $\langle I_{}\rangle\tau_{p}\ll 1$ (i.e.,
photocurrent pulses do not overlap). For example, for a detector
with a time constant $\tau_p=10$ ns, the corresponding photon flux must be
below
$10^{8}$ photons/s. In terms of power, for the wavelength of $500$ nm,
it means about 10 pW.

Eqs. (\ref{self-corr-curr2}) and (\ref{cross- corr-curr2}) become
then
\begin{eqnarray}\label{self-corr-curr-klyshko}
\langle i_{1}(t) i_{1}(t+\tau)\rangle= \eta_{1}\langle
q_{1}^{2}\rangle \mathcal{F}(\tau) \langle I_{1}\rangle,
\end{eqnarray}
\begin{eqnarray}\label{cross- corr-curr2-klyshko}
\langle i_{1}(t)i_{2}(t+\tau)\rangle=\eta_{1}\eta_{2}\langle
q_{1}\rangle\langle q_{2}\rangle \mathcal{F}(\tau) \langle
I_{1}\rangle.
\end{eqnarray}

The same equations has been found in \cite{klysh} and the quantum
efficiency has been estimated as
\begin{equation}\label{eta-klyshko}
  \Gamma_{2}=\eta_{2}\langle q_{2}\rangle=\frac{\langle
q_{1}^{2}\rangle}{\langle q_{1}\rangle^{2}}\langle q_{1}\rangle
  \frac{\langle i_{1}(t)i_{2}(t+\tau)\rangle}{\langle i_{1}(t)
  i_{1}(t+\tau)\rangle}.
\end{equation}
This formula is not satisfying from the metrological point of view
because of the presence of some unknown parameter related to the
statistics of charge fluctuations that we discussed previously and
that has to be estimated in some other way. We suggest to avoid
the problem by integrating Eq. (\ref{cross- corr-curr2-klyshko})
over time $\tau$. It could be done after the acquisition of the
profile of the function has been performed. By definition
(\ref{F(t)}), it is evident that $\int d\tau\mathcal{F}(\tau)=1$;
integrating Eq. (\ref{cross- corr-curr2-klyshko}) in $\tau$ and
dividing it by Eq. (\ref{curr-time}), we obtain
\begin{equation}\label{eta-klyshko-integ}
\Gamma_{2}=\eta_{2}\langle q_{2}\rangle=\frac{\int d\tau\langle
i_{1}(t)i_{2}(t+\tau)\rangle}{\langle i_{1}\rangle}
\end{equation}

As pointed out, another drawback of (\ref{eta-klyshko}) is the
necessity to work only at very low intensity, where no overlapping
between pulses happens. This, in terms of experiment, means that
one has to work in the so-called "charge accumulation mode", in
which the electron charge is accumulated until reaching some
detectable threshold. In the case of avalanche devices, it
corresponds to the possibility of direct photon-counting.

\textbf{(II)} $\langle I_{}\rangle\tau_{p}\gtrsim 1$ but still
$v^{2}\ll 1$ (i.e., photocurrent pulses overlap but the parametric
gain and photon flux are still quite low). Considering the same
parameters as used in case I, coherence time $\tau_{coh}$ of the
order of $100$ fs, and the requirement that $v^2\le0.001$, this
means a photon flux between $10^8$ photons/s and $10^10$ photons/s
or power between $10$ pW and $100$ nW.

Eqs. (\ref{self-corr-curr2}) and (\ref{cross- corr-curr2}) become
\begin{eqnarray}\label{self-corr-curr-mod}
\langle i_{1}(t) i_{1}(t+\tau)\rangle=\langle
i_{1}\rangle^{2}+\eta_{1}\langle q_{1}^{2}\rangle
\mathcal{F}(\tau)\langle I_{1}\rangle,
\end{eqnarray}
\begin{eqnarray}\label{cross-corr-curr-mod}
\langle i_{1}(t)i_{2}(t+\tau)\rangle=\langle i_{1}\rangle\langle
i_{2}\rangle+\eta_{1}\eta_{2}\langle q_{1}\rangle\langle
q_{2}\rangle \mathcal{F}(\tau) \langle I_{1}\rangle.
\end{eqnarray}

We stress that in \cite{klysh}, to the best of our knowledge the
only theoretical paper treating the absolute calibration of analog
detectors using PDC, the very low parametric gain was assumed from
the very beginning. In particular, all terms proportional to
$v^{4}$ were neglected. This corresponds to the approximation (I)
in our treatment. Accordingly, in~\cite{klysh} the limitations of
Eq. (\ref{eta-klyshko}) were not discussed. Our analysis shows
that when the intensity is high enough to yield a strongly analog
current, Eqs. (\ref{self-corr-curr-mod}) and
(\ref{cross-corr-curr-mod}) should be used. Now we define the
correlation functions of the current fluctuations as
\begin{eqnarray}\label{current fluctuation}
\langle \delta i_{k}(t)\delta i_{l}(t+\tau)\rangle&\equiv& \langle
i_{k}(t)i_{l}(t+\tau)\rangle -\nonumber\\
&-&\langle i_{k}(t)\rangle\langle
i_{l}(t+\tau)\rangle\quad(k,l=1,2)\:.\qquad.
\end{eqnarray}
We underline that these functions remain in principle
experimentally estimable. So a new formula, similar to
(\ref{eta-klyshko}), is available for analog quantum efficiency
estimation:
\begin{equation}\label{eta-klyshko-MI}
  \Gamma_{2}=\eta_{2}\langle q_{2}\rangle=\frac{\langle
q_{1}^{2}\rangle}{\langle q_{1}\rangle^{2}}\langle q_{1}\rangle
  \frac{\langle \delta i_{1}(t)\delta i_{2}(t+\tau)\rangle}{\langle \delta
i_{1}(t) \delta
  i_{1}(t+\tau)\rangle}.
\end{equation}
Once again, the drawback of this formula is the presence of the
unknown parameter $\langle q_{1}^{2}\rangle/\langle
q_{1}\rangle^{2}$ that requires additional measurements to be
performed. As before, integrating in $\tau$ the expression for the
cross-correlation, i.e., the definition (\ref{current
fluctuation}) for $k=1,l=2$, we obtain
\begin{equation}\label{eta-klyshko-integ-MI}
\Gamma_{2}=\eta_{2}\langle q_{2}\rangle=\frac{\int d\tau\langle \delta
i_{1}(t)\delta i_{2}(t+\tau)\rangle}{\langle
i_{1}\rangle}\:.
\end{equation}
This equation represents one of the main results of the paper,
since it shows that the absolute calibration of analog detectors
by using SPDC is indeed possible.

A  drawback of quantum efficiency measurement in this regime could
derive from the fact that the terms proportional to the square of
intensity in (\ref{self-corr-curr-mod}) and
(\ref{cross-corr-curr-mod}) become rather large, much larger than
the term proportional to the intensity - the one that provides
calibration. We need to subtract this background, as it is done in
(\ref{current fluctuation}). Although it is an easily estimable
quantity, a little relative uncertainty could generate a large
uncertainty in the efficiency estimation: a limitative result when
looking toward metrological applications. The physical reason for
this behaviour can be found recalling that the quantum correlation
so attractive in PDC has the scale of $\tau_{coh}$. In the analog
regime this correlation is almost deleted because of the averaging
over a time $\tau_{p}\gg\tau_{coh}$. Anyway, as long as we take
$\langle I_{}\rangle\tau_{p}\sim 1\div10$, such a problem does not
arise.

An alternative could be working in the pulsed
regime, in which the duration of any pump pulses is not so far
from the coherence time of PDC and the distance between them is
larger than $\tau_{p}$. Of course, to have a large number of
photons during $\tau_{p}$ (strongly overlapping regime) we need to
increase the parametric gain. A detailed study of this possibility
will be presented elsewhere.

\textbf{(III)} $v^{2}\gtrsim 1$ (i.e., high-intensity regime).

In this regime each term of (\ref{self-corr-curr2}) and
(\ref{cross- corr-curr2}) is important and no general way can be
found for the absolute  calibration of analog detectors, at least
with a CW pump. It can be shown that in the single-mode case the
integral terms in (\ref{self-corr-curr2}) and (\ref{cross-
corr-curr2}) become equal to the first ones, proportional to the
square intensity, and thus could be easily estimated. In
the case of CW pump, single-mode detection requires very narrow filters and
fast detectors, beyond realistic present technological
possibilities. Furthermore, calculation shows that for analog
calibration we need to know the transmission spectrum of the
filters. Anyway, also in this case one can consider the
possibility to reach single-mode detection by using a pulsed pump.

\section{Squeezing for calibration} \vskip 0.5cm

In the 1980's and the 1990's a lot of work has been done to demonstrate
both theoretically and experimentally the possibility to obtain
sub-shot-noise  photocurrent statistics taking advantage of strong PDC
quantum
correlation. The shot-noise level (SNL) is defined to be the lower
limit to the photocurrent noise level, which is achieved for coherent states
of the
field. Basically, for what concerns two-mode
squeezing, two different kinds of schemes have been used. The
first one consists of detecting
the currents from the two light beams (signal and idler) and
subtracting them \cite{Smithey}. The variance of the difference
current or the difference between the numbers of generated
electrons can be less than the same quantity measured for coherent
beams. In the other scheme the information about fluctuations in
one beam is used to manipulate the intensity of the second beam
(feedforward technique) \cite{Mertz}, or directly the pump
intensity (feedback technique) \cite{Tapster} in order to correct
the fluctuations. The goal in this case is getting a reduction of
photocurrent fluctuations below the shot-noise level for the detector
measuring one
of the beams.

It is important that in both schemes the minimum reachable
squeezing factor depends strongly on the quantum efficiency of the
detectors. Thus, it is reasonable to consider the possibility to
extract the quantum efficiency from the degree of squeezing.

Let us first consider the two-mode squeezing introducing the
difference between the currents $i_{-}=i_{1}-i_{2}$. Using
equations (\ref{self-corr-curr2}) and (\ref{cross- corr-curr2})
the autocorrelation function can be easily evaluated. When we
consider two balanced detectors, $\eta_{1}\langle
q_{1}\rangle=\eta_{2}\langle q_{2}\rangle=\eta\langle q\rangle$,
collecting photons from symmetric and equal detection areas, and
no fluctuations in the charge produced per photon occur, i.e.,
$\langle q^{2}\rangle=\langle q\rangle^{2}$, we have
\begin{equation}\label{variance}
  \langle\delta i_{-}(t)\delta i_{-}(t+\tau) \rangle=SNL(1-\eta),
\end{equation}
where the shot noise level is given by $\hbox{SNL}=2\langle
q\rangle^{2}\eta\langle I\rangle\mathcal{F}(\tau)$. Let us discuss
this formula. First of all, its validity does not disappear in the
high-gain regime and at high light intensity because the integral
terms of (\ref{self-corr-curr2}) and (\ref{cross- corr-curr2})
cancel each other in the calculation of (\ref{variance}). Quite
surprisingly, it shows that some aspects of the quantum
correlation, such as the possibility of shot-noise suppression,
can be preserved when working with macroscopic intensities, where
the cross-correlation is dominated by classical terms proportional
to the square of the intensity. Furthermore, it opens the
possibility of calibrating analog detectors with SPDC at high
parametric gain. In particular, it can be useful in the
calibration of detectors in which the electronic noise of the
external amplifier is dominant compared to the shot noise when
working at low intensities (cases (I) and (II) of the previous
section). In this kind of detectors, like, for instance, simple
photodiodes, no avalanche or multiplication occurs and in general
we can consider the condition $\langle q^{2}\rangle=\langle
q\rangle^{2}$ satisfied. So the requirements for the derivation of
(\ref{variance}) are realistic in this case. It can be used for
calibration directly, by calibrating the SNL with the help of a
coherent source producing the same average current as measured for
one of the SPDC beams, or by integrating in $\tau$ and then
normalizing by $\langle i_{1}\rangle$:
\begin{equation}\label{two mode sq}
\frac{\int d\tau\langle\delta i_{-}(t)\delta i_{-}(t+\tau)
\rangle}{\langle i_{1}\rangle}=2\langle q\rangle(1-\eta).
\end{equation}

From the experimental viewpoint, the validity of (\ref{variance})
is guaranteed if the detection volumes of the two detectors are
conjugated. In other words, detector 2 has to collect exactly all
and only the modes conjugated with those collected by detector 1.
For that pertaining the spatial modes, detailed realistic
calculation performed in \cite{BGBL} as well as recent experiments
\cite{PRL932004} show that the detection areas should be large
enough to collect several spatial modes. By satisfying this
condition, one would also increase the detected radiation power
and hence, the signal-to-noise ratio.

Finally, we note that an equivalent formula is valid for the
counting regime without any assumptions except
$\eta_{1}=\eta_{2}=\eta$:
\begin{equation}\label{two mode sq N}
\frac{\langle(\delta N_{-})^{2}\rangle}{\langle
N\rangle}=2(1-\eta),
\end{equation}
where $N_{-}=N_{1}-N_{2}$ is the difference between the numbers of
counts in detectors 1 and 2, and $\langle(\delta
N_{-})^{2}\rangle$ is its variance. This method of calibration
could be interesting because, unlike the method of coincidence
counting based on Eq. (\ref{3}), it only requires counting
photodetection pulses.

Of course the assumption of balanced quantum efficiencies used for
deriving (\ref{variance})  somehow complicates the aim of reaching
the accuracy needed for metrological applications, although the
technique of balancing detectors is largely used in quantum
optics.

Finally, we consider the possibility to exploit the feedback or
feedforward schemes. In Ref \cite{Mertz}, the latter is studied in
detail for optical parametric oscillator (OPO) above threshold
where one has bright average intensity component and small
fluctuations. It is shown that the minimum of fluctuations
achievable can be expressed in terms of the frequency spectra of
the fluctuations in each beam and the correlation spectrum in the
absence of the feedforward action. Unfortunately the results of
\cite{Mertz} can not be applied directly to SPDC because in the
case of SPDC the resulting state is squeezed vacuum where high
fluctuations of intensity occur. Anyway, description of the
feedforward scheme presented in \cite{Mertz} can be applied to the
currents. For instance let the current in the detector registering
beam 1, $i_{1}$, be varied by using a modulation mechanism
conditioned by the measurement performed on beam 2. If
$\widetilde{ i_{1}}$ is the current registered after modulation,
in our time representation (for simplicity with $\tau=0$) the result of
Ref. \cite{Mertz} yields
\begin{equation}\label{feedforward}
\langle (\delta\widetilde{ i_{1}})^{2}\rangle=
\langle (\delta i_{1})^{2}\rangle-\frac{\langle \delta i_{1} \delta
i_{2}\rangle^{2}}{\langle (\delta
i_{2})
^{2}\rangle}\:.
\end{equation}
This formula shows that is possible to measure the reduction in
the current fluctuations of detector 2 after modulation instead of
measuring the cross-correlation function $\langle \delta i_{1}
\delta i_{2}\rangle$. Substituting (\ref{self-corr-curr-mod}) and
(\ref{cross-corr-curr-mod}) into (\ref{feedforward}), we have
\begin{equation}\label{tapster}
\frac{\langle (\delta\widetilde{ i_{1}})^{2}\rangle}{\langle
(\delta i_{1})^{2}\rangle}= 1-\eta_{1}\eta_{2}\frac{\langle
q_{1}\rangle^{2}\langle q_{2}\rangle^{2}}{\langle
q_{1}^{2}\rangle\langle q_{2}^{2}\rangle}\:.
\end{equation}
If the two detectors have balanced quantum efficiencies and excess
charge noise is absent, the above formula can be used for the
efficiency estimation, representing a further interesting option
in this sense.

It is interesting to observe that in \cite{Tapster}, the authors
obtained the same  suppression  of fluctuations as given by
(\ref{tapster}) using a theoretical model describing a feedback
procedure applied to SPDC. The only difference is that our
equation takes into account the statistics of the charge, which
enters through the factor $\frac{\langle q_{1}\rangle^{2}\langle
q_{2}\rangle^{2}}{\langle q_{1}^{2}\rangle\langle
q_{2}^{2}\rangle}$. Even if the quantum efficiency is ideal, this
factor prevents one from reaching perfect noise reduction.

\section{  Conclusion} \vskip 0.5cm

Motivated by the necessity of a general absolute calibration
scheme for analog detectors for various applications and aiming to
extend the absolute calibration scheme from the single-photon to
the analog regime, we have performed a systematical study on the
possibility of applying SPDC calibration methods to the analog
regime. Possibilities and limitations following from the specific
properties of SPDC have been investigated.

Our results show that measurement of the correlations in the
output currents indeed can be used to extend the absolute
calibration method to the analog regime, although the experimental
implementation will require an accurate study and solution of some
technical problems.

In particular, it is shown that integration of the photocurrent
correlation functions in time allows one to avoid the measurement of
the photocurrent pulse shape and to eliminate the necessity to know
the statistics of electrons in the photocurrent.

Also, our analysis showed the possibility to go beyond the regime of
non-overlapping
photocurrent pulses, which was used in earlier works, and to pass to higher
intensities. A possible way to reduce
the background in the measured photocurrent correlation function, which will
unavoidably accompany the transition to higher intensities, namely, passing
to
the pulsed regime of PDC, is outlined.

Finally, we studied the new subject of squeezing as a tool for
absolute photodetectors calibration. In particular, two-mode
squeezing is shown to be the only way of performing calibration in
the high-intensity regime of SPDC, where other methods fail, but
with the limitation that one should have two detectors with
balanced quantum efficiencies. Also, a possibility of estimating
the quantum efficiency for the feedforward or feedback scheme of
fluctuations suppression is considered, reaching analogous conclusions.

\section{Acknowledgements}
The Turin group acknowledges  the support of MIUR
(FIRB RBAU01L5AZ-002) and Fondazione San Paolo. The Russian group
acknowledges the support of Russian Foundation for Basic Research
(grant \# 05-02-1639). One of the authors, I.R., thanks Moscow State
University for kind hospitality during the realization of this work.


\end{document}